\documentclass[conference]{IEEEtran}
\usepackage{graphicx}
\usepackage{epstopdf}
\usepackage{fancyhdr}
\usepackage{amsmath,amssymb,amstext}
\usepackage{subeqnarray}
\usepackage{cite}
\usepackage{hhline}
\usepackage{bm}
\usepackage{soul}
\usepackage{times,amsmath,amsfonts}
\usepackage{subfigure}
\usepackage{amsmath,amssymb}
\usepackage{graphicx}
\usepackage{cite}
\usepackage{subfigure}
\usepackage{subeqnarray}
\usepackage{color}

\usepackage{multirow}
\usepackage{booktabs}
\usepackage{hyperref}
\usepackage{algorithmic}
\usepackage[ruled, linesnumbered]{algorithm2e}
\ifCLASSINFOpdf
\else
   \usepackage{graphicx}
   \graphicspath{{../eps/}}
   \DeclareGraphicsExtensions{.eps}
\fi
\definecolor{r}{rgb}{1,0,0}
\definecolor{b}{rgb}{0,0,1}
\definecolor{k}{rgb}{0,1,1}

\newcounter{saveeqn}%

\allowdisplaybreaks
\DeclareMathSymbol{\Phi}{\mathord}{letters}{8}

\begin{document}
\title{Ray Antenna Array Enhanced Low-Altitude ISAC: Performance Analysis and Beamforming Design}

\author{
\IEEEauthorblockN{$\text{Zhiqiang~Xiao}^*$, $\text{Tao~Zhang}^*$, $\text{Hao~Wu}^*$, $\text{Xiaoqiang~Qiao}^*$, $\text{Jiang~Zhang}^*$, $\text{Zhenjun~Dong}^{\dag\S}$, and $\text{Yong~Zeng}^{\dag\S}$
}
\IEEEauthorblockA{*The Sixty-third Research Institute, National University of Defense Technology, Nanjing 210007, China\\
$\dag$National Mobile Communications Research Laboratory, Southeast University, Nanjing 210096, China\\
$\S$Purple Mountain Laboratories, Nanjing 211111, China\\
Email: \{xzqnudt, qxq0527, god2525775\}@163.com, \{whao1983, ztcool\}@126.com, \{zhenjun\_dong, yong\_zeng\}@seu.edu.cn
}
}
\maketitle

\begin{abstract}
The low-altitude economy (LAE) heavily relies on aerial vehicles, yet these platforms remain vulnerable to environmental and security risks, necessitating robust airspace monitoring.
Integrated sensing and communication (ISAC) as one of the key technologies of 6G provides potential solutions for safe LAE.
However, conventional antenna arrays face limitations in cost, scalability, and coverage, especially directly above the base station, due to hardware complexity and degraded angular resolution.
By exploiting the recently proposed ray antenna array (RAA), this paper considers a RAA-enhanced low-altitude ISAC system.
RAA architecture employs multiple ray-arranged arrays directly connected without phase shifters, significantly reducing hardware costs while supporting flexible beamforming via dynamic ray selection.
Moreover, RAA can provide uniform angular resolution and eliminates coverage holes, making it particularly suitable for low-altitude ISAC.
In this paper, we formulate an optimization problem for joint ray selection and beamforming to enhance sensing coverage under communication constraints.
An efficient alternating optimization algorithm is proposed to solve this problem.
Analytical and simulation results demonstrate that RAA achieves higher sensing signal-to-noise ratio compared to traditional arrays, offering a cost-effective and high-performance solution for achieving low-altitude ISAC.
\end{abstract}

\section{Introduction}
Low-altitude economy (LAE) has attracted increasing attentions recently, which represents a comprehensive system capable of supporting a wide range of low-altitude (generally below 1000 meters) applications, such as express delivery, aerial transportation, and three-dimensional (3D) environment mapping \cite{jiang20236g,jiang20236g2}.
However, both manned and unmanned aerial vehicles (UAVs), serving as essential aerial platforms for LAE, are highly vulnerable to environment disturbances, adverse weather conditions, and malicious interventions, which may cause flight accidents, such as collisions, crashes, or airspace trespassing.
This underscores the urgent needs of enhancing low-altitude airspace monitoring \cite{song2025overview}.

As one of the key technologies for six-generation (6G) networks, integrated sensing and communication (ISAC) demonstrates significant potentials for supporting the LAE.
By utilizing existing densely deployed base stations (BSs), 6G aims to not only provide communications for conventional ground users but also enable ubiquitous sensing and communications \cite{zhang2020perceptive}.
To this end, prior works mainly focus on exploring the sensing capabilities of a single BS, with many efforts devoted for waveform design \cite{xiao2022waveform}, beamforming optimization \cite{liu2021cramer}, and prototyping experiments \cite{luo2025isac}, etc.
However, its sensing performance is constrained by practical issues such as severe path loss, signal obstruction, and fluctuations of the target's radar cross section (RCS).
Recently, cooperative or networked ISAC has drawn growing research interest \cite{li2023toward,meng2024cooperative,cheng2025networked}.

However, the aforementioned literatures mainly focus on the traditional antenna architectures.
Such architectures face challenges for achieving the goals of low-altitude ISAC.
First, to further exploit spatial diversity and resolution for communication and sensing, larger array aperture is necessary for the next-generation transceivers.
However, traditional array architectures do not well scale with the array size.
Even for the hybrid analog/digitgal beamforming (HBF), where the number of radio frequency (RF) chains can be significantly reduced, the hardware cost of phase shifters is still prohibitive, especially in the era of extremely large-scale multiple-input multiple-ouput (XL-MIMO) \cite{lu2024tutorial}.
Moreover, the high resolution phase shifter design is challenging, particularly in high frequency regimes like millimeter wave (mmWave) and Terahertz bands.
Second, traditional antenna arrays, like uniform linear array (ULA), suffer from the degraded angular resolution when the users/targets depart from the array boresight.
This may lead to the sensing coverage hole directly above the BS.

To address these challenges, researchers have begun to investigate various novel antenna array architectures for ISAC.
In \cite{xiu2025movable}, the authors investigated the movable antenna (MA)-enabled ISAC system, which outperforms the conventional fixed-antenna array, by jointly optimizing MAs' positions and beamforming.
In \cite{xiong2025intelligent}, a novel intelligent rotatable antenna (IRA) architecture is introduced, which enhances the signal strength and coverage by dynamically adjusting the antenna boresights.
Furthermore, \cite{adhikary2024holographic} proposed a holographic MIMO-based cell-free network that can significantly improve the energy efficiency of ISAC systems.
However, these advanced architectures often require additional controllers or specialized antennas, potentially introducing additional hardware complexity and implementation challenges.

In this paper, we explore the recently proposed ray antenna array (RAA) for low-altitude ISAC.
RAA consists of multiple ray-like arranged simple ULAs (sULAs), where antenna elements of each sULA are directly connected without requiring any phase shifter \cite{dong2025ray}.
Note that even without relying any analog or digital beamforming, each sULA can still form a beam whose mainlobe direction matches with its boresight \cite{dong2025ray2}.
Then, by dynamically selecting the appropriate sULAs to be connected with the RF chains for baseband processing, RAA can achieve a flexible beamforming.
Although RAA requires more antenna elements compared to the conventional arrays, no phase shifter is needed and the prices of antenna elements is quite low, thus RAA enjoys a significant hareware cost reduction \cite{dong2025ray}.
Moreover, RAA can achieve uniform angular resolution across all possible directions, without suffering from the angular resolution degradation as that of conventional ULA \cite{jiang2025ray}.
In addition, as each sULA only needs to cover a small angular range, antenna elements with higher direction can be applied for beamforming gain improvement.

In the following, we first briefly introduce RAA and describe the considered RAA-based low-altitude ISAC system.
After that, a joint ray selection and beamforming optimization problem is formulated, which aims to enhance the sensing coverage while guaranteeing the communication performance.
As such a problem is non-convex that is difficult to be directly solved,
to draw some insights, the sensing performance of RAA is first analyzed, where the maximum achievable sensing signal-to-noise ratio (SNR) considering the impact of the antenna pattern is derived.
After that, a novel alternating optimization algorithm is proposed to solve this problem.
Comparison analysis shows that RAA enables higher sensing SNR than conventional ULA, by employing the directional antenna elements.
Simulation results are provided to demonstrate the potentials of RAA for low-altitude ISAC.

\section{System Description and Channel Model}
\subsection{System Description and RAA Architecture}
\begin{figure} %system model Fig.1%
  \centering
  % Requires \usepackage{graphicx}
  \includegraphics[width=0.48\textwidth]{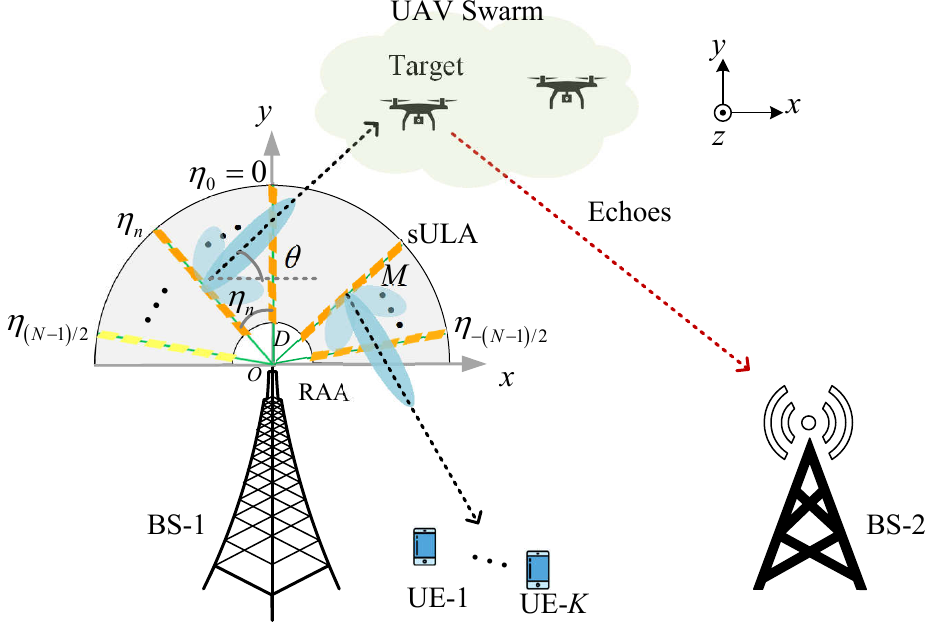}
  \caption{Illustration for the considered RAA-based low-altitude ISAC system.}\label{system}\vspace{-15pt}
\end{figure}

As illustrated in Fig.\ref{system}, we consider a RAA-based low-altitude ISAC system, where BS-1 equipped with the recently proposed RAA wishes to communicate with $K$ ground user equipments (UEs) while simultaneously sensing aerial targets by cooperating with BS-2.
The RAA consists of $N$ ray-like arranged sULAs, where each sULA has $M$ antenna elements that are directly connected without requiring any phase shifter.

The orientation of each sULA is deliberately designed. Specifically, for clarity of illustration, a Cartesian coordinate system is shown in Fig.~\ref{system}.
Here, all $N$ sULAs are symmetrically placed around the positive $y$-axis.
Denote by $\eta_n$ the orientation of the $n$th sULA and let the sULA of $n=0$ aligned with the positive $y$-axis as the reference, with $\eta_0=0$.
It has $\eta_n\in[-\eta_{\max},\eta_{\max}]$, $n\in\mathcal{N}$, where $0\le\eta_{\max}\le\pi/2$ and $\mathcal{N}=\{-\frac{N-1}{2},\cdots,\frac{N-1}{2}\}$ is the index set of all sULAs.
Here, we assume that $N$ is an odd number and $\eta_n$ at the left hand side of $y$-axis is positive for ease of exposition.
In addition, to avoid the mutual coupling among the antenna elements across sULAs, the distance from the origin $O$ to the first antenna element of each sULA is set as $D>0$ \cite{dong2025ray}.

Consider an arbitrary angle-of-departure (AoD) $\theta$ w.r.t. the positive $x$-axis
%\footnote{Note that for notational convenience, the AoD $\theta$ and orientation angle $\eta_n, n\in\mathcal{N}$ are defined based on different reference directions.}
, with $\theta\in[-\theta_{\max},\theta_{\max}]$ and $0\le\theta_{\max}\le\pi/2$.
Without loss of generality,  the adjacent antenna element spacing of each sULA is $d=\frac{\lambda}{2}$, with $\lambda$ being the signal wavelength.
The steering vector of the $n$th sULA for $\theta$ is
$
\mathbf{a}(\theta,\eta_n) = \left[1,e^{j\pi\sin(\theta-\eta_n)},\cdots,e^{j\pi(M-1)\sin(\theta-\eta_n)}\right]^T,
$
where $\theta-\eta_n$ is a relative angle between the AoD $\theta$ and the boresight of the $n$th sULA, satisfying that $0\le |\theta-\eta_n|\le \pi/2$.
The array response matrix for the AoD $\theta$ is expressed as
\begin{equation}\label{array response matrix}
  \mathbf{A}(\theta) = [\mathbf{a}(\theta,\eta_n)]_{n\in\mathcal{N}}\times\mathrm{diag}(\mathbf{b}(\theta))\in\mathbb{C}^{M\times N},
\end{equation}
where  $\mathbf{b}(\theta)\triangleq[b(\theta,\eta_n)]_{n\in\mathcal{N}}\in\mathbb{C}^{N\times1}$
with $b(\theta,\eta_n)$ being the antenna response of the reference antenna element of the $n$th sULA, which is given by
\begin{equation}\label{antenna response}
  b(\theta,\eta_n) = e^{j\frac{2\pi}{\lambda}D\sin(\theta-\eta_n)}\sqrt{G_n(\theta-\eta_n)},
\end{equation}
where $G_n(\cdot)$ denotes the transmit antenna element pattern of the $n$th sULA.
Therefore, without requiring any beamforming, the transmit array response of the $n$th sULA at $\theta$ is
\begin{equation}\label{transmit array response}
\begin{aligned}
  r(\theta,\eta_n) &= \sqrt{1/M}\mathbf{1}_{M\times1}^T\mathbf{a}(\theta,\eta_n)b(\theta,\eta_n)\\
  %&=\frac{b(\theta,\eta_n)}{\sqrt{M}}\sum\limits_{m=0}^{M-1}e^{j\pi m\sin(\theta-\eta_n)}\\
  &=e^{\frac{2\pi}{\lambda}D\sin(\theta-\eta_n)}\sqrt{MG_n(\theta-\eta_n)}\mathcal{H}_M(\sin(\theta-\eta_n)),
\end{aligned}
\end{equation}
where
$\mathcal{H}_M(x)\triangleq e^{j\frac{\pi}{2}(M-1)x}\frac{\sin(\frac{\pi}{2}Mx)}{M\sin(\frac{\pi}{2}x)}
$
is the Dirichlet function.
Here, it is assumed that the mainlobe beamwidth of the antenna pattern $G_n(\cdot)$ is wider than that of $\mathcal{H}_M(\cdot)$ and antenna gain $G_n(\theta-\eta_n)$ is symmetric and peaks at $\theta=\eta_n$.

{\it Remark 1}: Note that \eqref{transmit array response} demonstrates the {\it orientation-dependent energy-focusing} capability of the sULA without requiring any phase shifter.
Specifically, the transmit power of the sULA can be magnified by approximately $M$ times when $\theta-\eta_n\approx 0$.
It implies that after baseband signal processing, by selecting the appropriate sULAs to be connected with the RF chains, RAA can achieve a flexible beamforming as that of the traditional HBF systems.
However, different from the traditional HBF that usually requires large number of phase shifters, a ray selection network (RSN) consisting of low-cost switches is introduced\cite{dong2025ray}.
Therefore, although the RAA requires more antenna elements (i.e., $MN$), it may achieve a significant hardware cost saving since no phase shifter is needed and antenna elements are much cheaper.

{\it Remark 2}:
%As shown in \eqref{transmit array response}, the mainlobe beamwidth of the array response is determined by $\mathcal{H}_M(\sin(\theta-\eta_n))$.
%The null points of $\mathcal{H}_M(\sin(\theta-\eta_n))$ can be obtained by letting $\frac{\pi}{2}M\sin(\theta-\eta_n)=\pm p\pi, p=1,2,\cdots$.
%Thus, the null-to-null mainlobe beamwidth of $\mathcal{H}_M(\sin(\theta-\eta_n))$ can be derived as $\theta_{BW}=2\arcsin(2/M)$
The angular resolution of RAA corresponding to the half of mainlobe beamwidth is given by $\triangle\theta=\arcsin(2/M)$ \cite{jiang2025ray}.
Note that it is independent of the irradiation direction $\theta$, which implies that RAA can achieve the {\it uniform angular resolution} across all directions \cite{jiang2025ray}.
%However, as $\arcsin(\cdot)$ is nonlinear, the null points of the array response for each sULA is unevenly spaced, i.e., $\arcsin(2n/M)\neq n\times\arcsin(2/M)$.
To effective suppress the interference between adjacent sULAs, the orientation angles of all $N$ sULAs are designed as $\eta_n = n\times\arcsin({2}/{M}), n\in\mathcal{N}$.
The number of required sULAs is  $N=2\lfloor\frac{\eta_{\max}}{\arcsin(2/M)}\rfloor+1$.
%Note that for $\eta_{\max}=\pi/2$, it has $N\approx\lfloor\pi M/2\rfloor$ when $M\gg 1$.
Moreover, to ensure that the antenna spacing across all sULAs is no smaller than $\lambda/2$, the distance $D$ should satisfy that $D\ge\lambda/(4\sin(0.5\arcsin(2/M)))$ \cite{dong2025ray2}.

{\it Remark 3}:
As shown in \eqref{antenna response} and \eqref{transmit array response}, the total angular coverage range $[-\theta_{\max},\theta_{\max}]$ is naturally divided by RAA consisting of $N$ sULAs with different orientation angles.
Each sULA is only responsible for a small portion of the total coverage.
Thus, RAA offers a new design degree-of-freedom (DoF) to configure the antenna pattern $G_n(\cdot)$ for each sULA~$n$ to satisfy different performance requirements.
%, which refers to as {\it antenna pattern diversity}.
%In particular, antenna elements with stronger directivity can be utilized for enhancing the beamforming gain.

These above advantages of RAA make it particularly appealing for the emerging low-altitude ISAC scenarios, where the aerial targets may easily locate directly above the BS, which cannot be effectively covered with traditional antenna array architecture.
This thus motivates our current work.

%\begin{figure*} %system model Fig.1%
%  \centering
%  % Requires \usepackage{graphicx}
%  \subfigure[ULA]{
%  \includegraphics[width=0.38\textwidth]{HBF.pdf}\label{collocated_system}
%  }
%  \subfigure[RAA]{
%  \includegraphics[width=0.42\textwidth]{RAA.pdf}\label{seperated_system}
%  }
%  \caption{Comparison of traditional ULA and proposed RAA for HBF.}\label{RAA_vs_HBF}\vspace{-10pt}
%\end{figure*}

\subsection{RAA-based ISAC Channel Model}
In this paper, we consider a common time-frequency resource block used by BS-1 to downlink communicate with $K$ single-antenna UE, while as the transmit signals can be prior known by another BS via backhaul, they can serve as the bi-static radar to sense aerial targets.
The system is assumed to be operated at the mmWave frequency band.

Let $\mathcal{K}=\{1,\cdots,K\}$ denote the index set of UEs.
The downlink communication channel between BS-1 equipped with RAA and the $k$th single-antenna UE is
\begin{equation}\label{downlink communication}
\begin{aligned}
  \mathbf{h}_k &= \left[\mathbf{h}^T_{k,1},\cdots,\mathbf{h}^T_{k,N}\right]^T=\sum\nolimits_{l=1}^{L_k}\alpha_{k,l}\mathrm{vec}\left(\mathbf{A}(\theta_{k,l})\right), k\in\mathcal{K},
\end{aligned}
\end{equation}
where $\mathbf{h}_{k,n}\triangleq\sum\nolimits_{l=1}^{L_k}{\alpha}_{k,l}\mathbf{a}(\theta_{k,l},\eta_n)$ , $L_k$ is the number of multi-paths, and $\alpha_{k,l}$ denotes the path loss of the $l$th path.
On the other hand, for bi-static sensing, as this paper focuses on the RAA design at the transmitter side, it assumes that BS-2 only employs a single antenna.
The bi-static sensing channel between BS-1 and BS-2 is
\begin{equation}\label{radar channel}
  \mathbf{h}_s = \sum\nolimits_{q=1}^{Q}\beta_q\mathrm{vec}\left(\mathbf{A}(\theta_q)\right),
\end{equation}
where $Q$ denotes the number of aerial targets, $\beta_q$ represents two-way path loss associated with the $q$th target, and $\theta_q$ is the AoD of the $q$th target.

Denote by $\mathbf{s}\triangleq[s_1,\cdots,s_K]^T\in\mathbb{C}^{K\times1}$ the transmit information-bearing symbol for $K$ UEs, with  $\mathbb{E}[\mathbf{s}\mathbf{s}^H]=\mathbf{I}_K$.
The digital transmit beamforming matrix is represented as $\mathbf{W}=[\mathbf{w}_1,\cdots,\mathbf{w}_K]\in\mathbb{C}^{N_{\mathrm{RF}}\times K}$, where $\mathbf{w}_k\in\mathbb{C}^{N_{\mathrm{RF}}\times1}$ denotes the transmit beamforming vector for the $k$th user, with $N_{\mathrm{RF}}$ being the number of RF chains.
Here, it considers that $K\le N_{\mathrm{RF}}<N$.
The baseband processing output is $\mathbf{x}=\mathbf{W}\mathbf{s}$.
Then, the RSN is introduced to dynamically select $N_{\mathrm{RF}}$ out of $N$ sULAs to be connected with the RF chains for transmission \cite{dong2025ray}.
The selection matrix is defined as $\mathbf{U}\in\mathbb{R}^{N_{\mathrm{RF}}\times N}$, which is a binary matrix, satisfying that $\left\|[\mathbf{U}]_{i,:}\right\|_0=1$ and $\left\|[\mathbf{U}]_{:,j}\right\|_0\le1$ for $i=1,\cdots,N_{\mathrm{RF}}$ and $j=1,\cdots,N$.
After that, the output at $N$ ports denoted by $\bar{\mathbf{x}}\in\mathbb{C}^{N\times1}$ can be expressed as
\begin{equation}
\bar{\mathbf{x}} = \mathbf{U}^H\mathbf{W}\mathbf{s} = \mathbf{U}^H\sum\nolimits_{k=1}^{K}\mathbf{w}_ks_k.
\end{equation}
The transmit power of $\bar{\mathbf{x}}$ satisfies that
\begin{equation}\label{transmit power}
  \mathbb{E}[\|\bar{\mathbf{x}}\|^2]=\mathbb{E}[\|\mathbf{U}^H\mathbf{W}\mathbf{s}\|^2]\overset{(a)}{=}\sum\nolimits_{k=1}^{K}\|\mathbf{w}_k\|^2\le P_t,
\end{equation}
where $P_t$ denotes the transmit power constraint and $(a)$ holds due to $\mathbf{U}\mathbf{U}^H=\mathbf{I}_{N_\mathrm{RF}}$ and $\mathbb{E}[\mathbf{s}\mathbf{s}^H]=\mathbf{I}_{K}$.
Then, a power splitter is applied to connect with $M$ antennas for each sULA.
Therefore, the transmit signal of RAA for all $MN$ antenna element is
$\mathbf{x}_t = \bar{\mathbf{x}}\otimes\sqrt{1/M}\mathbf{1}_{M\times1}.$

With downlink communication channel in \eqref{downlink communication}, the received signal at the $k$th UE is
\begin{equation}
\begin{aligned}
  y_k &= \mathbf{h}_k^H\mathbf{x}_t + z_k\\
%      &=\underbrace{\left(\sum\nolimits_{l=1}^{L_k}\alpha_{k,l}\mathbf{r}(\theta_{k,l})\right)^H}_{\bar{\mathbf{h}}_k^H}\bar{\mathbf{x}}+z_k\\
      &=\bar{\mathbf{h}}_k^H\mathbf{U}^H\sum\nolimits_{k=1}^K\mathbf{w}_ks_k+z_k, k\in\mathcal{K},
\end{aligned}
\end{equation}
where $\bar{\mathbf{h}}_k\triangleq\left(\sum\nolimits_{l=1}^{L_k}\alpha_{k,l}\mathbf{r}(\theta_{k,l})\right)\in\mathbb{C}^{N\times1}$ denotes the effective channel between $N$ sULA ports and the $k$th UE, with $\mathbf{r}(\theta_{k,l})\triangleq~\left[r(\theta_k,\eta_1),\cdots,r(\theta_k,\eta_N)\right]^T$ according to \eqref{transmit array response}, and $z_k\sim\mathcal{CN}(0,\sigma^2)$ is the additive white Gaussian noise (AWGN).
Therefore, for the $k$th UE, the received signal-to-interference-plus-noise ratio (SINR) is
\begin{equation}\label{communication SINR}
\gamma_k(\mathbf{U},\mathbf{W}) = \frac{\left|\bar{\mathbf{h}}_k^H\mathbf{U}^H\mathbf{w}_k\right|^2}{\sum\limits_{i\neq k}\left|\bar{\mathbf{h}}_k^H\mathbf{U}^H\mathbf{w}_i\right|^2+\sigma^2}, k\in\mathcal{K}.
\end{equation}

On the other hand, with the bi-static sensing channel in \eqref{radar channel}, the received echoes at BS-2 is
\begin{equation}
y_s = \bar{\mathbf{h}}_s^H\mathbf{U}^H\sum\nolimits_{k=1}^K\mathbf{w}_ks_k + z_s,
\end{equation}
where $\bar{\mathbf{h}}_s\triangleq\sum\nolimits_{q=1}^{Q}\beta_q\mathbf{r}(\theta_q)$  denotes the effective sensing channel of the RAA and $z_s\sim\mathcal{CN}(0,\sigma_s^2)$ is the AWGN.
Here, to highlight the potential of RAA for sensing, we focus on the single-target case.
Thus, the received sensing SNR for the AoD $\theta$ can be derived as
\begin{equation}\label{sensing SNR}
  \gamma_s(\mathbf{U},\mathbf{W})= \sum\nolimits_{k=1}^K{\big|\beta\mathbf{r}(\theta)^H\mathbf{U}^H\mathbf{w}_k\big|^2}/{\sigma_s^2}.
\end{equation}
Note that such design can be extended to multi-target sensing by successively removing the contributions of estimated targets, which is deferred as our  future work.

\section{Performance Analysis and Optimization}
From \eqref{communication SINR} and \eqref{sensing SNR} it observes that both communication and sensing performance critically depend on the ray selection matrix $\mathbf{U}$ and beamforming matrix $\mathbf{W}$.
Therefore, a joint ray selection and beamforming optimization problem is considered to maximize the minimum communication SINR in \eqref{communication SINR}, subject to the ray selection and transmit power constraints, while guaranteeing that the sensing SNR in \eqref{sensing SNR} is no smaller than a threshold $\gamma_{th}$, as follows
\begin{align}
\mathrm{(P1):}
& \max\limits_{\mathbf{U},\mathbf{W}}\ \min\limits_{k\in\mathcal{K}}\gamma_k \label{P1}\\
& \text{s.t.}  \quad [\mathbf{U}]_{i,j}\in\{0,1\}, \forall i, j,  \tag{\ref{P1}{a}}\label{U_c1}\\
&     \qquad  \|[\mathbf{U}]_{i,:}\|_0=1, \forall i, \tag{\ref{P1}{b}}\label{U_c2}\\
&     \qquad  \|[\mathbf{U}]_{:,j}\|_0\le 1,\forall j,\tag{\ref{P1}{c}}\label{U_c3}\\
&     \qquad   \sum\nolimits_{k=1}^{K}\|\mathbf{w}_k\|^2\le P_t, \tag{\ref{P1}{d}}\label{transmit_power_constraint}\\
&      \qquad   \gamma_s\ge\gamma_{th}\tag{\ref{P1}{e}}\label{sensing constraint}.
\end{align}
Note that due to the binary constraints on $\mathbf{U}$ and the tight coupling of $\mathbf{U}$ and $\mathbf{W}$, it is difficult to directly solve $(\mathrm{P1})$.
In the following, we first analyze the RAA-based sensing performance to draw some essential insights.

\subsection{RAA-based Sensing Performance Analysis}
%\begin{figure} %system model Fig.1%
%  \centering
%  % Requires \usepackage{graphicx}
%  \includegraphics[width=0.48\textwidth]{RAA_orientation.eps}
%  \caption{Illustration for the best and worst case of RAA-based sensing.}\label{resolution}\vspace{-0.3cm}
%\end{figure}
To maximize the sensing SNR for a single target without considering the communication requirements, only one RF chain is sufficient.
Thus, the problem becomes to select the sULA $n^*\in\mathcal{N}$ whose boresight is closest to the AoD $\theta$.
According to \eqref{sensing SNR}, the resulted maximum sensing SNR is
\begin{equation}\label{maximum sensing SNR}
\gamma_{\text{RAA},\max}(\theta) = |\beta|^2\bar{P}_tMG_{\text{RAA}}(\theta-\eta_{n^*})|\mathcal{H}_M(\sin(\theta-\eta_{n^*}))|^2,
\end{equation}
where $\bar{P}_t\triangleq P_t/\sigma_s^2$ and here we consider that all antenna elements of RAA have a common antenna pattern $G_{\text{RAA}}(\cdot)$ for convenience.
Note that the sensing SNR depends on the AoD $\theta$ associated with the target location.
If $\theta$ is aligned with the boresight of the $n^*$th sULA, i.e., $\theta-\eta_{n^*}=0$, $n^*\in\mathcal{N}$, sULA $n^*$ can focus the transmit power at $\theta$, which refers to as the best case.
According to \eqref{maximum sensing SNR}, it has
\begin{equation}\label{best}
\gamma_{\text{RAA},\max}^{\text{best}} = |\beta|^2\bar{P}_tMG_{\text{RAA}}(0).
\end{equation}
In contrast, the worst case happens when $|\theta-\eta_{n^*}|=\frac{1}{2}\triangle\theta$, $n^*\in\mathcal{N}$.
The resulted sensing SNR for the worst case is
\begin{equation}\label{worst}
\begin{aligned}
\gamma_{\text{RAA},\max}^{\text{worst}} =
|\beta|^2\bar{P}_tMG_{\text{RAA}}(\triangle\theta/2)|\mathcal{H}_M(\sin(\triangle\theta/2))|^2.
\end{aligned}
\end{equation}
To guarantee the sensing coverage, the sensing SNR for the worst case should be no smaller than $\gamma_{th}$, i.e., $\gamma_{\text{RAA},\max}^{\text{worst}}\ge \gamma_{th}$.
Thus, the antenna element pattern should satisfy that
$G_{\text{RAA}}(\triangle\theta/2)\ge\frac{\gamma_{th}}{\xi|\mathcal{H}_M(\sin(\triangle\theta/2))|^2}$, with $\xi\triangleq|\beta|^2\bar{P}_tM$ and the 3dB beamwidth of $G_{\text{RAA}}(\cdot)$ denoted by $\theta_{\text{3dB,RAA}}$ should be no smaller than the RAA angular resolution, i.e., $\theta_{\text{3dB,RAA}}\ge \triangle\theta=\arcsin(2/M)$.

\subsection{Joint Ray Selection and Beamforming Optimization}
One common approach to solve (P1) is to alternately optimize $\mathbf{U}$ and $\mathbf{W}$.
For the fixed $\mathbf{U}$, by introducing an auxiliary variable $\bar{\gamma}$, the problem (P1) reduces to
\begin{equation}
\begin{aligned}
(\mathrm{P2.1}): &\max\limits_{\mathbf{W},\bar{\gamma}}\ \bar{\gamma}\\
&\text{s.t.}\ \ \gamma_{k}\ge \bar{\gamma},  k\in\mathcal{K},\\
&\qquad \eqref{transmit_power_constraint} \ \text{and} \ \eqref{sensing constraint},
\end{aligned}
\end{equation}
which is still non-convex due to the quadratic and fractional terms in $\gamma_s$ and $\gamma_k$.
According to \eqref{communication SINR}, the communication constraint can be reformulated as the convex second-order cone (SOC) constraints as
\begin{equation}\label{soc2}
\begin{aligned}
&\|\bar{\mathbf{h}}_k^H\mathbf{U}^H\mathbf{w}_1,\cdots,\bar{\mathbf{h}}_k^H\mathbf{U}^H\mathbf{w}_K,\sigma\|\le\sqrt{1+\frac{1}{\bar{\gamma}}}\mathrm{Re}\left(\bar{\mathbf{h}}_k^H\mathbf{U}^H\mathbf{w}_k\right),\\
&\mathrm{Im}\left(\bar{\mathbf{h}}_k^H\mathbf{U}^H\mathbf{w}_k\right)=0, k\in\mathcal{K}.
\end{aligned}
\end{equation}

%According to \eqref{communication SINR}, the communication constraint can be equivalently expressed as
%\begin{equation}\label{reformed communication constraint}
%\sum\nolimits_{i=1}^K\left|\bar{\mathbf{h}}_k^H\mathbf{U}^H\mathbf{w}_i\right|^2 +\sigma^2 \le \left(1+\frac{1}{\bar{\gamma}}\right)\left|\bar{\mathbf{h}}_k^H\mathbf{U}^H\mathbf{w}_k\right|^2.
%\end{equation}
%Here, as the phase rotations of the optimal $\{\mathbf{w}_k\}_{k=1}^K$ do not alter the inequality constraint \eqref{reformed communication constraint}, we can impose that $\mathrm{Im}(\bar{\mathbf{h}}_k^H\mathbf{U}^H\mathbf{w}_k)=0$, $k\in\mathcal{K}$.
%Thus, \eqref{reformed communication constraint} can be reformulated as the convex second-order cone (SOC) constraints as
%\begin{equation}\label{soc2}
%\begin{aligned}
%&\|\bar{\mathbf{h}}_k^H\mathbf{U}^H\mathbf{w}_1,\cdots,\bar{\mathbf{h}}_k^H\mathbf{U}^H\mathbf{w}_K,\sigma\|\le\sqrt{1+\frac{1}{\bar{\gamma}}}\mathrm{Re}\left(\bar{\mathbf{h}}_k^H\mathbf{U}^H\mathbf{w}_k\right),\\
%&\mathrm{Im}\left(\bar{\mathbf{h}}_k^H\mathbf{U}^H\mathbf{w}_k\right)=0, k\in\mathcal{K}.
%\end{aligned}
%\end{equation}
\begin{algorithm}
	\caption{Alternating Optimization for Solving Problem (P1)}
	\label{alg1}
	\textbf{Input}: $\bar{\mathbf{h}}_k$, $k\in\mathcal{K}$, $\theta$, $\tilde{\gamma}_{th}$, $P_t$.

    \textbf{Output}: $\mathbf{W}^*$, $\mathbf{U}^*$, $\bar{\gamma}^*$, $\gamma_s^*$.

    Initialize feasible $\mathbf{W}^{(0)}$, $\mathbf{U}^{(u)}$, $\bar{\gamma}^{(0)}$, $u=0$, $\epsilon_1$, $\epsilon_2$.

	\Repeat{$|\bar{\gamma}_{u+1}-\bar{\gamma}_{u}|\le \epsilon_2$}
    {
        Fix $\mathbf{U}=\mathbf{U}^{(u)}$; Initialize $i=0$;

        \Repeat{$|\bar{\gamma}^{(i+1)}-\bar{\gamma}^{(i)}|\le \epsilon_1$}
        {
            Solve the problem (P2.1.1) under the given $\mathbf{W}^{(i)}$, to get the optimal $\mathbf{W}^{*}$ and $\bar{\gamma}^{*}$;

            Updating $\mathbf{W}^{(i+1)}=\mathbf{W}^{*}$;

            Updating $\bar{\gamma}^{(i+1)} = \bar{\gamma}^*$;

            $i = i + 1$;
        }

        $\bar{\gamma}_u=\bar{\gamma}^*$;

        Fix $\mathbf{W}^*$, solve the problem (P2.2) via exhaustive searching to get the optimal $\mathbf{U}^*$ and $\bar{\gamma}^*$.

        Updating $\bar{\gamma}_{u+1}=\bar{\gamma}^*$;

        Updating $\mathbf{U}^{(u+1)} = \mathbf{U}^*$;

        $u = u + 1$;
    }	
\end{algorithm}
However, the non-convex sensing constraint in \eqref{sensing constraint} cannot be transformed into a SOC constraint.
Here, we apply the successive convex approximation (SCA) technique to deal with the non-convex \eqref{sensing constraint}.
Denoted by $g\left(\mathbf{W}\right)\triangleq\sum\nolimits_{k=1}^K|\mathbf{r}_{\mathrm{U}}(\theta)^H\mathbf{w}_k|^2$ with $\mathbf{r}_{\mathrm{U}}(\theta)\triangleq\mathbf{U}\mathbf{r}(\theta)\in\mathbb{C}^{N_{\text{RF}}\times1}$, \eqref{sensing constraint} can be equivalently expressed as
$g(\mathbf{W})\ge\frac{\gamma_{th}\sigma_s^2}{|\beta|^2}\triangleq\tilde{\gamma}_{th}$.
As $g(\mathbf{W})$ is a convex function, it can be lower bounded by its first-order Taylor expansion as
\begin{equation}\label{sca}
\small
\begin{aligned}
g(\mathbf{W})&\ge  g(\mathbf{W}^{(i)})+2\sum\limits_{k=1}^K\mathrm{Re}\left(\mathbf{w}_k^{(i)H}\mathbf{r}_{\mathrm{U}}(\theta)\mathbf{r}_{\mathrm{U}}(\theta)^H(\mathbf{w}_k-\mathbf{w}_k^{(i)})\right)\\
&\triangleq \varphi_{i}(\mathbf{W},\mathbf{W}^{(i)}),
\end{aligned}
\end{equation}
where $\mathbf{W}^{(i)}\triangleq[\mathbf{w}_1^{(i)},\cdots,\mathbf{w}_K^{(i)}]$ is a given local point and $\varphi_{i}(\mathbf{W},\mathbf{W}^{(i)})$ is the lower bound function of $g(\mathbf{W})$ at the $i$th iteration.
Note that $g(\mathbf{W})$ and $\varphi_{i}(\mathbf{W},\mathbf{W}^{(i)})$
have the identical gradient.
Then, by replacing the communication constraint and \eqref{sensing constraint} with \eqref{soc2} and \eqref{sca}, respectively, the problem (P2.1) can be recast as the following convex problem
\begin{equation}
\begin{aligned}
\mathrm{(P2.1.1):}
& \max\limits_{\mathbf{W},\bar{\gamma}}\ \bar{\gamma}\\
& \text{s.t.}  \quad %\|\bar{\mathbf{h}}_k^H\mathbf{U}^H\mathbf{w}_1,\cdots,\bar{\mathbf{h}}_k^H\mathbf{U}^H\mathbf{w}_K,\sigma\|\notag\\
%&\qquad\qquad\le\sqrt{1+\frac{1}{\bar{\gamma}}}\mathrm{Re}\left(\bar{\mathbf{h}}_k^H\mathbf{U}^H\mathbf{w}_k\right),
%k\in\mathcal{K},  \tag{\ref{P2}{a}}\\
\eqref{soc2}, \eqref{transmit_power_constraint}, \\
%&\qquad\ \mathrm{Im}\left(\bar{\mathbf{h}}_k^H\mathbf{U}^H\mathbf{w}_k\right)=0, \tag{\ref{P2}{b}} k\in\mathcal{K},\\
%&     \qquad   \sum\nolimits_{k=1}^{K}\|\mathbf{w}_k\|^2\le P_t, \tag{\ref{P2}{c}}\\
&\qquad \varphi_i(\mathbf{W},\mathbf{W}^{(i)})\ge\tilde{\gamma}_{th},
\end{aligned}
\end{equation}
which can be efficiently solved by applying the standard convex optimization techniques or readily toolbox, like CVX.
Then, by iteratively updating the local point with $\mathbf{W}^{(i+1)}=\mathbf{W}^{*}$ and solving a sequence of convex optimization problem (P2.1.1), the problem (P2.1) can be effectively solved.

Next, with the obtained feasible $\mathbf{W}^*=[\mathbf{w}_1^*,\cdots,\mathbf{w}_K^*]$, the original problem (P1) reduces to
\begin{equation}
\begin{aligned}
(\mathrm{P2.2}): &\max\limits_{\mathbf{U}}\min\limits_{k\in\mathcal{K}} \gamma_k \\
&\text{s.t.} \quad \eqref{U_c1}-\eqref{U_c3}\ \text{and} \ \eqref{sensing constraint},
\end{aligned}
\end{equation}
which is a combinatorial search problem.
By exhaustive searching all feasible solutions satisfying the constraints \eqref{U_c1}-\eqref{U_c3} and \eqref{sensing constraint}, the optimal solution of (P2.2) can be obtained as $\mathbf{U}^*$.
Then, by alternatively optimizing $\mathbf{W}$ and $\mathbf{U}$ by solving (P2.1) and (P2.2) until $\bar{\gamma}^*$ converges, the problem (P1) can be solved, which is summarized in Algorithm 1.

\section{RAA versus ULA for Sensing}
In this section, we compare RAA with the conventional ULA in terms of sensing performance, while their communication performance has been discussed in \cite{dong2025ray2}.
Consider the ULA consisting of $M$ antenna elements aligned with the positive $y$-axis, i.e., $\eta_0=0$.
The steering vector of the ULA for $\theta$ is
$\mathbf{a}_{\text{ULA}}(\theta) = \left[1, e^{-j\pi\sin\theta},\cdots,e^{-j\pi(M-1)\sin\theta}\right]^T$.
For the conventional discrete Fourier transform (DFT) codebook for analog beamforming, the codeword is designed as $\mathbf{f}_n=\frac{1}{\sqrt{M}}[1,e^{-j\pi\frac{2n}{M}},\cdots,e^{-j\pi(M-1)\frac{2n}{M}}]^T$, $n\in\mathcal{N}'\triangleq\{-\frac{N'-1}{2},\cdots,\frac{N'-1}{2}\}$, where  $N'=2\lfloor\frac{M\sin\theta_{\max}}{2}\rfloor+1$ is the number of codewords.
To maximize the sensing SNR without considering the communication constraints, the analog beamformer should match with the steering vector $\mathbf{a}(\theta)$.
Therefore, the resulted maximum sensing SNR including the impact of antenna element pattern is
\begin{equation}
\begin{aligned}
\gamma_{\text{ULA},\max}(\theta)&=|\beta|^2P_tG_{\text{ULA}}(\theta)\left|\mathbf{a}^H(\theta)\mathbf{f}_{n^*}\right|^2/\sigma_s^2\\
&=|\beta|^2\bar{P_t}MG_{\text{ULA}}(\theta)\left|\mathcal{H}_M(\sin\theta-2n^*/M)\right|^2,
\end{aligned}
\end{equation}
where $\mathbf{f}_{n^*}$ corresponds to the codeword whose direction is closest to $\theta$ and $G_{\text{ULA}}(\cdot)$ denotes the antenna element pattern of ULA.
For the best case, i.e., $\sin\theta=\frac{2n^*}{M}$, $n^*\in\mathcal{N}'$, the resulted sensing SNR is
\begin{equation}\label{ULA_best}
\gamma_{\text{ULA},\max}^{\text{best}}(\theta) = |\beta|^2\bar{P}_tMG_{\text{ULA}}(\theta).
\end{equation}
On the other hand, similar to \eqref{worst}, when $|\sin\theta-\frac{2n^*}{M}|=\frac{1}{M}$, $n^*\in\mathcal{N}'$, the sensing SNR for the worst case is derived as
\begin{equation}
\gamma_{\text{ULA},\max}^{\text{worst}}(\theta) =
|\beta|^2\bar{P}_tM{G}_{\text{ULA}}(\theta)|\mathcal{H}_M(1/M)|^2.
\end{equation}

Note that different from that of RAA in \eqref{maximum sensing SNR} and \eqref{best}, the boresight of ULA is always aligned with the positive $x$-axis and the antenna element pattern $G_{\text{ULA}}(\theta)$ peaks at $\theta=0$.
To guarantee the sensing coverage performance, i.e., $\gamma_{s,\text{ULA}}^{\text{worst}}(\theta)\ge\gamma_{th}$, the antenna element pattern of ULA satisfies that $G_{\text{ULA}}(\theta)\ge\frac{\gamma_{th}}{\xi|\mathcal{H}_M(1/M)|^2}$ and the 3dB beamwidth of ULA antenna element should cover the whole angular range of interest, i.e., $\theta_{\text{3dB,ULA}}\ge 2\theta_{\max}$.
Given the sum of antenna gain $G_{\text{sum}}$ and the 3dB beamwidth $\theta_{\text{3dB}}$, the peak antenna gain can be expressed as $G(0)\approx{G_{\text{sum}}}/({\epsilon\theta_{\text{3dB}}})$, with $\epsilon$ is a positive constant \cite{dong2025ray2}.
As the required 3dB beamwidth of ULA antenna is much wider than that of RAA, with the fixed $G_{\text{sum}}$, it has
\begin{equation}\label{antenna ratio}
\frac{G_{\text{RAA}}(0)}{G_\text{ULA}(0)}=\frac{\theta_{\text{3dB,ULA}}}{\theta_{\text{3dB,RAA}}}\gg1,
\end{equation}
which implies that the RAA employed the antenna elements with narrower beamwidth can achieve higher beamforming gain.
From \eqref{worst} and \eqref{ULA_best}, we obtain that as long as
\begin{equation} G_{\text{RAA}}\left(\frac{\theta_{{\text{3dB,RAA}}}}{2}\right)=\frac{1}{2}G_{\text{RAA}}(0)\ge\frac{G_{\text{ULA}}(0)}{|\mathcal{H}_M(\sin(\triangle\theta/2))|^2},
\end{equation}
it has $\gamma_{\text{RAA},\max}^{\text{worst}}\ge\gamma_{\text{ULA},\max}^{\text{best}}(\theta)$, $\theta\in[-\theta_{\max},\theta_{\max})$.
It means that the sensing SNR of RAA is always greater than that of ULA for better sensing coverage.

%\begin{figure*}[htbp]
%    \centering
%        \subfigure[The convergence of the proposed Algorithm 1.]{
%    \includegraphics[width=0.33\textwidth]{covergence.eps}\label{RAA dir}
%    }
%    \hspace{-15pt}
%    \subfigure[$\bar{\gamma}$ versus $\gamma_{th}$ for RAA-based ISAC.]{
%    \includegraphics[width=0.33\textwidth]{RAA_ISAC2.eps}\label{RAA_ISAC}
%    }
%    \hspace{-15pt}
%    \subfigure[Transmit beampattern of RAA for ISAC.]{
%    \includegraphics[width=0.332\textwidth]{RAA_ISAC_beampattern.eps}\label{RAA_ISAC_beampattern}
%    }
%  \caption{Illustration of the performance of RAA-based ISAC.}\label{RAA_ISAC}\vspace{-10pt}
%\end{figure*}

%\begin{figure*}[htbp]
%    \centering
%        \subfigure[The convergence of the proposed Algorithm 1.]{
%    \includegraphics[width=0.33\textwidth]{covergence.eps}\label{RAA dir}
%    }
%    \hspace{-15pt}
%    \subfigure[$\bar{\gamma}$ versus $\gamma_{th}$ for RAA-based ISAC.]{
%    \includegraphics[width=0.33\textwidth]{RAA_ISAC2.eps}\label{RAA_ISAC}
%    }
%    \hspace{-15pt}
%    \subfigure[Transmit beampattern of RAA for ISAC.]{
%    \includegraphics[width=0.332\textwidth]{RAA_ISAC_beampattern.eps}\label{RAA_ISAC_beampattern}
%    }
%  \caption{Illustration of the performance of RAA-based ISAC.}\label{RAA_ISAC}\vspace{-10pt}
%\end{figure*}
\begin{figure*}[htbp]
    \centering
        \subfigure[RAA with directional antenna]{
    \includegraphics[width=0.33\textwidth]{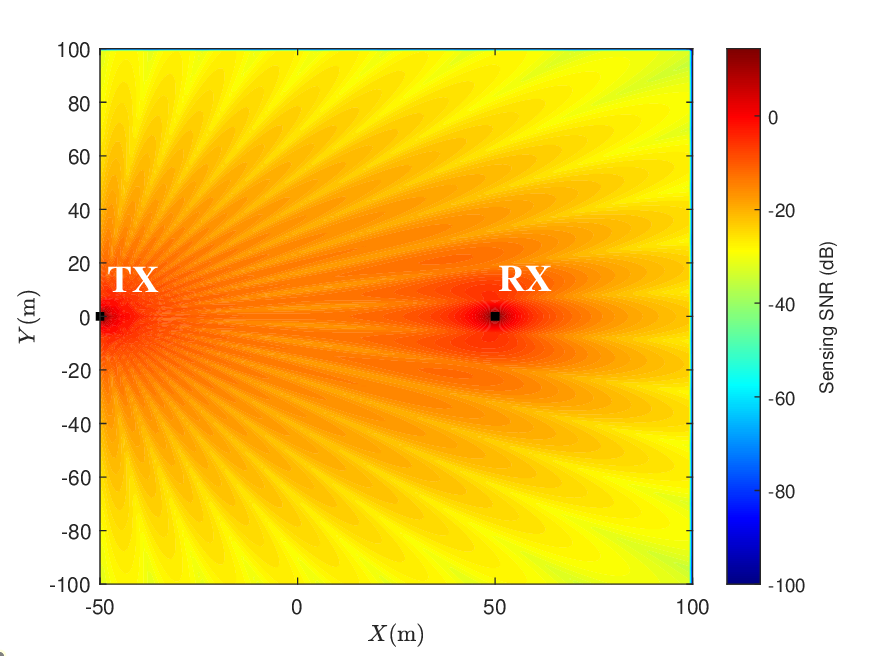}\label{RAA dir}
    }
    \hspace{-15pt}
    \subfigure[ULA with omnidirectional antenna]{
    \includegraphics[width=0.33\textwidth]{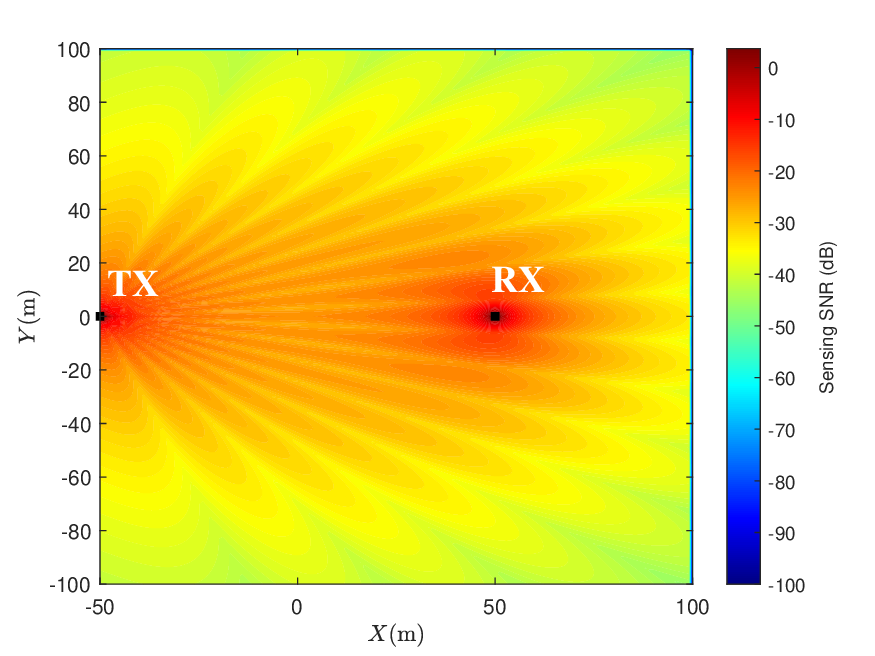}\label{ULA omn}
    }
    \hspace{-15pt}
    \subfigure[ULA with directional antenna]{
    \includegraphics[width=0.332\textwidth]{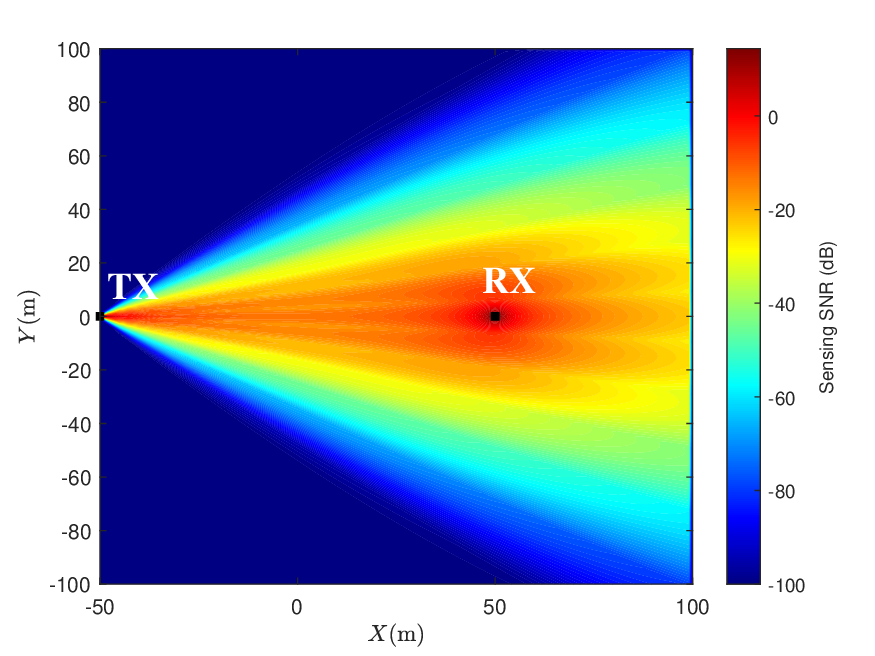}\label{ULA dir}
    }
  \caption{Comparison of RAA and conventional ULA in terms of sensing SNR.}\label{SenSNR}\vspace{-16pt}
\end{figure*}
\begin{figure*}[htbp]
    \centering
        \subfigure[The convergence of the proposed Algorithm 1.]{
    \includegraphics[width=0.33\textwidth]{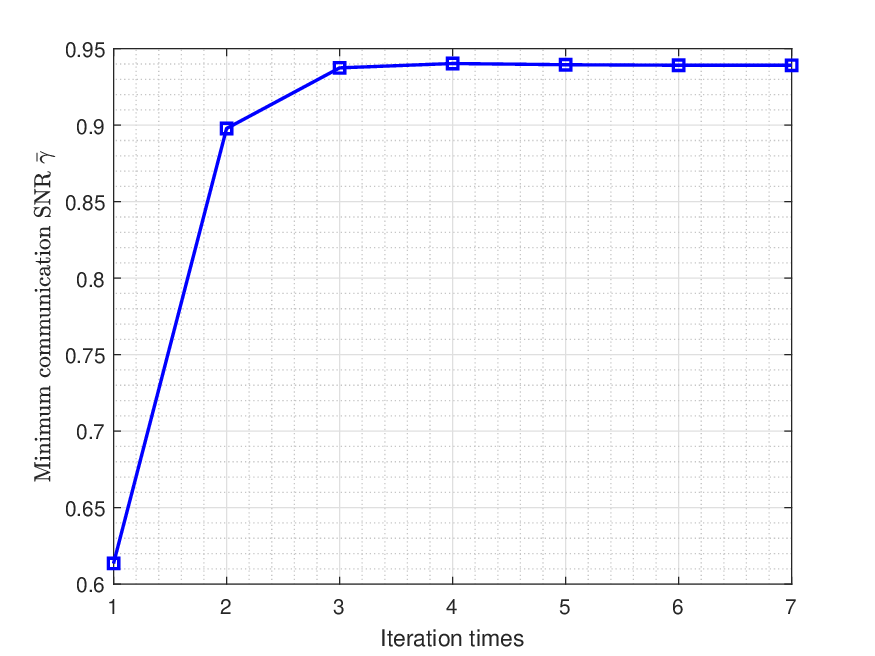}\label{RAA dir}
    }
    \hspace{-15pt}
    \subfigure[Comparison of RAA and ULA for ISAC]{
    \includegraphics[width=0.33\textwidth]{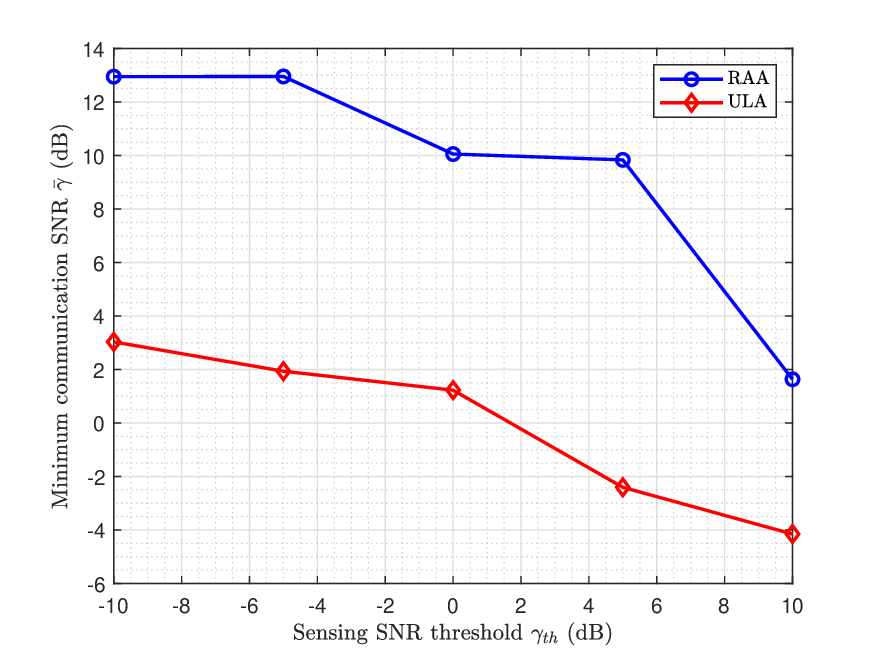}\label{RAA_ISAC}
    }
    \hspace{-15pt}
    \subfigure[Transmit beampattern of RAA for ISAC.]{
    \includegraphics[width=0.332\textwidth]{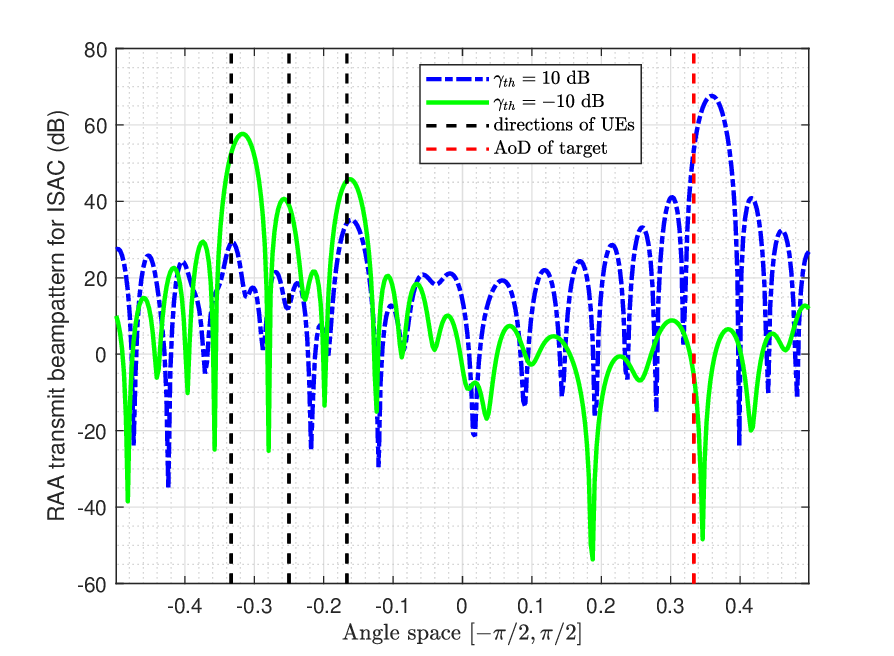}\label{RAA_ISAC_beampattern}
    }
  \caption{The performance of RAA for ISAC.}\label{RAA_ISAC}\vspace{-15pt}
\end{figure*}
\section{Simulation Results}
In this section, simulation results are provided to evaluate the potentials of RAA for low-altitude ISAC.
The number antenna elements for each sULA and RF chains is set as $M=16$ and $N_{\text{RF}}=3$, respectively.
To cover $\theta\in[-\pi/2,\pi/2)$, the maximum orientation of sULAs is set as $\eta_{\max}=\pi/2$, and $N=2\lfloor\frac{\eta_{\max}}{\arcsin(2/M)}\rfloor+1=25$.
%The number of sULAs and RF chains are $N=2\lfloor\frac{\eta_{\max}}{\arcsin(2/M)}\rfloor+1=25$ and $N_{\text{RF}}=3$, respectively.
The number of antenna elements for the conventional ULA is $M=16$.
%The DFT codebook is applied  and the number of codewords is $N'=2\lfloor\frac{M\sin\theta_{\max}}{2}\rfloor +1 = 17$.
Here, a Gaussian antenna pattern is considered for both RAA and ULA antenna elements, which is given by $G(\theta)=G(0)e^{-a\theta^2/\theta_{\text{3dB}}^2}$, where $a$ is a constant to control the beamwidth of the mainlobe and typically it has $a\approx2.7726$\cite{balanis2016antenna}.
For conventional ULA, the 3dB beamwidth is set as $\theta_{3dB}^{\text{ULA}}=\pi$ and the antenna pattern for ULA is
$G_{\text{ULA}}(\theta) = G_{\text{ULA}}(0)e^{-a\theta^2/\theta_{3dB,\text{ULA}}^2}$, with $G_{\text{ULA}}(0) = 17$ dB.
%By contrast, the 3dB beamwidth of RAA antenna elements only needs to be greater than its angular resolution, i.e., $\theta_{\text{3dB,RAA}}\ge\triangle\theta=\arcsin(2/M) = 0.0022\ \mathrm{rad}$.
Here, the 3dB beamwidth of RAA antenna elements is set as $\theta_{\text{3dB,RAA}}=\pi/12$.
Moreover, consider the sum of antenna gain $G_{\text{sum}}$ is same for both RAA and ULA, according to \eqref{antenna ratio}, it has $G_{\text{RAA}}(0)=\frac{G_{\text{ULA}}(0)\theta_{\text{3dB, ULA}}}{\theta_{\text{3dB,RAA}}}$.
Therefore, according to \eqref{antenna response}, the antenna pattern for the $n$th sULA of RAA is
$
  G(\theta-\eta_n) = \frac{G_{\text{ULA}}(0)\theta_{\text{3dB,ULA}}}{\theta_{\text{3dB,RAA}}}e^{-a(\theta-\eta_n)^2/\theta_{\text{3dB,RAA}}^2}.
$

Fig.~\ref{SenSNR} compares the resulted sensing SNRs of RAA and conventional ULA with omnidirectional and directional antenna, respectively.
Here, it considers that two BSs served as the bi-static radar to sense the target in a rectangular region with size of $150\mathrm{m}\times 100\mathrm{m}$, where the position of the target is denoted by $\mathbf{p}=[x,y,0]^{T}$, $-50\mathrm{m}<x<100\mathrm{m}$, $-50\mathrm{m}<y<50\mathrm{m}$.
It can be observed that the RAA achieves better sensing coverage than conventional ULA, especially in the edge regions where the latter suffers from the degraded sensing SNR, since these directions  deviate from the boresight of ULA, leading to reduced beamforming gain and lower SNR.
%The positions of two BSs are set as $\mathbf{b}_1=[-50\mathrm{m}, 0\mathrm{m}, 0\mathrm{m}]^T$ and $\mathbf{b}_2=[50\mathrm{m}, 0\mathrm{m}, 0\mathrm{m}]^T$.
%The distances from BS-1 to target and from target to BS-2 are $d_1=\|\mathbf{p}-\mathbf{b}_1\|$ and $d_2=\|\mathbf{p}-\mathbf{b}_2\|$, respectively.
%Then, the two-way propagation path loss can be obtained as $\beta=\frac{G_r\lambda^2\delta}{(4\pi)^3d_1^2d_2^2}$, where $\lambda=\frac{c}{f_c}$ is the signal wavelength, with $c=3\times10^8$~m/s and $f_c=28$ GHz being the speed of light and carrier frequency, respectively, $\delta=1\ \mathrm{m}^2$ is the RCS of target, and $G_r=17$ dB denotes the receive antenna gain of the omnidirectional antenna that is applied at BS-2.
%The transmit power is $P_t=1$ W and the noise power is $\sigma_s^2=-90$ dBm.
%In Fig.~\ref{SenSNR}, the RAA with directional antenna is compared to the ULA with omnidirectional antenna and directional antenna, respectively, in terms of the sensing SNR.
%It is observed that RAA achieves better sensing coverage than conventional ULA, particularly at the edge region, which cannot be well served by the conventional ULA, with the degraded sensing SNR, since the antenna element with higher directionality can be used for enhancing the beamforming gain.

Fig.~\ref{RAA_ISAC}(a) depicts the convergence of the proposed Algorithm 1 to solve the problem (P1) for RAA-based ISAC.
%It is observed that the maximized minimum communication SNR $\bar{\gamma}$ converges fast with a few iterations, which demonstrates the convergence and effectiveness of the proposed algorithm.
Fig.~\ref{RAA_ISAC}(b) compares RAA and ULA for ISAC, where the maximized minimum communication SINR $\bar{\gamma}$ decreases as the sensing SNR threshold $\gamma_{th}$ increasing for both RAA and ULA.
This is expected as $\gamma_{th}$ increases, more transmit power is focused on the direction of target, resulting the reduction of $\bar{\gamma}$, as shown in Fig.~\ref{RAA_ISAC}(c).
However, due to the beamforming gain brought by the directional antenna element of RAA, it can achieve better performance than that of ULA with omnidirectional antenna.

\section{Conclusion}
This paper studied low-altitude ISAC with the novel RAA architecture, where the communication SINR and sensing SNR of RAA are first derived.
Then, a joint ray selection and beamforming optimization problem for RAA-based ISAC was formulated.
To solve this non-convex optimization problem, a novel alternatively optimization algorithm was proposed.
Analytic and simulation results demonstrated that RAA outperforms traditional ULA for low-altitude ISAC by leveraging directional antenna pattern.

%\section*{Acknowledgment}
%This work was supported by the Natural Science Foundation
%of China under Grant 62371463, by the Innovation Research Foundation of National University of Defense Technology.

\bibliographystyle{IEEEtran}
\bibliography{RAA_ISAC_ref}

\end{document}